# Light (anti)nuclei production in high-energy nuclear collisions at the LHC with ALICE

**Chiara Pinto**[a,b,*] **on behalf of the ALICE Collaboration**

[a] *Department of Physics and Astronomy "E. Majorana", University of Catania,*
*Via S. Sofia 64, Catania, Italy*
[b] *INFN, Section of Catania,*
*Via S. Sofia 64, Catania, Italy*

*E-mail:* chiara.pinto@cern.ch

The measurement of (anti)nuclei production in pp, p-A and A-A collisions at ultrarelativistic energies is important to understand hadronization. The excellent tracking and particle identification capabilities of ALICE make it the most suited detector at the LHC to study light (anti)nuclei produced in high-energy hadronic collisions. (Anti)nuclei with mass numbers up to 4, such as (anti)deuterons, (anti)tritons, (anti)$^3$He and (anti)$^4$He have been successfully identified in ALICE at midrapidity ($|\eta|$<0.9). In this contribution, multiplicity dependent results on the yields, nuclei-to-protons ratios as well as the coalescence parameter $B_A$ as a function of the charged-particle multiplicity are presented and compared with the expectations of coalescence and statistical hadronization models (SHM) to provide insight into their production mechanism in heavy-ion collisions.

[*]Speaker

### 1. Physics motivation

The measurement of light (anti)nuclei produced in high-energy hadronic collisions provides information about their production mechanism. Their production mechanism is one of the open questions in heavy-ion physics because the binding energies of the produced nuclei (a few MeV) are low compared to the temperature of the system at the kinetic freeze-out ($T_{kin} \sim 10^2$ MeV) in which they are formed. Two classes of models are available to describe the measured production yields: the statistical hadronization and the coalescence models. In the Statistical Hadronization Model (SHM) [1], the hadrons are produced by a thermally equilibrated source and their abundances are fixed at the chemical freeze-out. On the other hand, the production of light (anti)nuclei can be explained via the coalescence of protons and neutrons which are close by in phase space at the kinetic freeze-out and match the spin, thus forming a nucleus [2]. The key parameter of the coalescence models is the coalescence parameter, which is related to the production probability of the nucleus via this process and can be calculated from the overlap of the nucleus wave function and the phase space distribution of the constituents via the Wigner formalism [3].

### 2. Analysis method

The production spectra of (anti)deuterons, (anti)$^3$H, (anti) $^3$He and (anti)$^4$He in pp and/or p-A collisions at several energies have been studied with ALICE. The ALICE apparatus is constituted by detectors placed at central rapidities (referred to as central barrel) and forward detectors. For the (anti)nuclei identification, the detectors of the central barrel ($|\eta|<0.9$) are used: the Inner Tracking System (ITS), the Time Projection Chamber (TPC) and the Time of Flight (TOF) detectors. The TPC specific energy loss signal (dE/dx) allows the separations of nuclei with Z=2 from the other charged particles produced in the collision, in the full momentum range. The same technique can only be used at low momenta for Z=1 nuclei (e.g. $p_T < 1.4$ GeV/c for deuterons), but using also the TOF it is possible to identify (anti)deuterons up to higher $p_T$.

Light nuclei measurements are affected by the contamination of secondary nuclei produced in the interaction of primary particles with the beam pipe and the detector material. This contamination is exponentially decreasing with increasing momentum. Most of the secondary particles from material have a large distance of closest approach (DCA) to the primary vertex and hence this item of information is used to correct for the contamination. Moreover, the residual secondary background is estimated and removed by means of Monte Carlo templates, which allow the discrimination between primary nuclei (with a DCA distribution peaked at zero) and secondary nuclei emitted isotropically (thus with a flat DCA distribution).

### 3. Results and discussion

In order to extract the light (anti)nuclei integrated yields, extrapolation to the unmeasured region has been performed by fitting the transverse momentum spectra with several functional forms, as the Blast Wave function [4] or the Lévy-Tsallis one [5]. As an example, the results for transverse momentum spectra of deuterons in p-Pb collisions at 8.16 TeV in several multiplicity classes are shown in Fig. 1 (left panel) and fitted with a Lévy-Tsallis function. The $p_T$ distributions become harder as the multiplicity increases: the spectra become less steep with increasing multiplicity and the mean transverse momentum moves towards higher values. Antiparticle-to-particle ratios are also measured in ALICE in several collision systems and energies. Such ratios resulted to be compatible with unity within uncertainties in all the studied systems, as expected in case of vanishing baryochemical potential at the LHC (see as an example the bottom-right plots in Fig. 1 for $^3$H and $^3$He ratios). The ratio between the measured yields of nuclei and that of protons is sensitive to the light nuclei production mechanism. In Fig. 2 the yield ratio to protons for deuterons (left panel), $^3$H and $^3$He (right panel) as a function of $<dN_{ch}/d\eta_{lab}>$ measured in pp, p-Pb and Pb-Pb collisions [6]-[9] is shown and compared to the expectations of the models. A smooth increase of this ratio with the system size is observed, reaching a constant value in Pb-Pb collisions. The two ratios show a similar trend with $<dN_{ch}/d\eta_{lab}>$, however the increase from pp to Pb-Pb results is about a factor of 3 larger for $^3$He/p than for d/p. The observed evolution of the d/p ratio is well described by the coalescence approach because of the increasing phase space in Pb-Pb. For high charged-particle multiplicity densities, the coalescence calculations and the canonical statistical model (CSM) expectations are close and both describe the behaviour of the data, within the current uncertainties. On the other hand, the models struggle to describe the ratio to protons for nuclei with A=3, as it is clear in the right panel of Fig. 2.

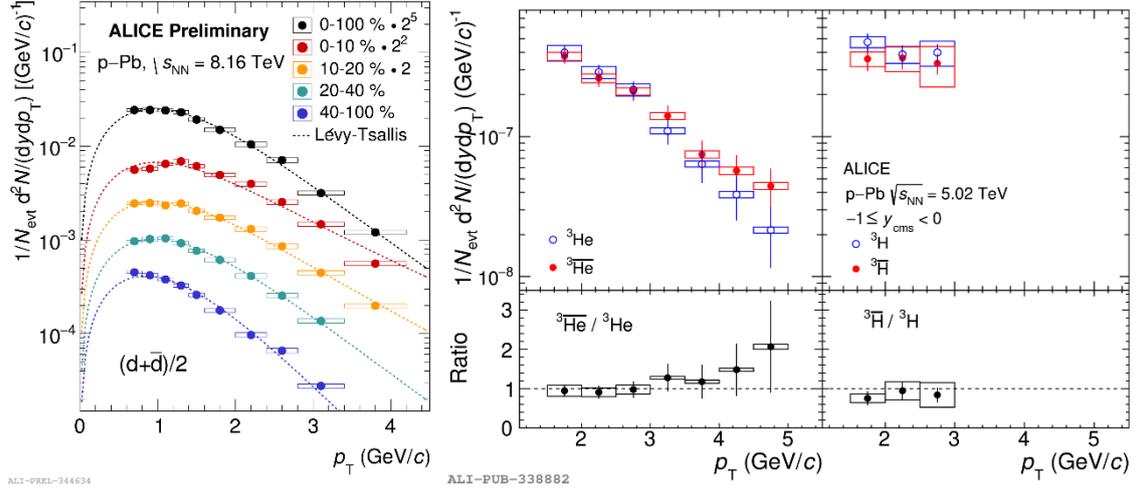

**Figure 1**: *Left.* Deuteron transverse momentum spectra in several multiplicity classes, fitted using a Lévy-Tsallis functional form in order to extrapolate to the unmeasured region. *Right.* (Anti)$^3$He and (anti)$^3$H transverse momentum spectra (top panels) as well as the antiparticle-to-particle ratios (bottom panels).

Fig. 3 shows the coalescence parameter $B_2$ as a function of the charged-particle multiplicity, for a fixed value of $p_T/A$. The measurements show a smooth transition from low charged-particle multiplicity densities, which refer to a small system size, to larger ones. The decreasing trend of $B_2$ with increasing $<dN_{ch}/d\eta_{lab}>$ suggests that the production mechanism in small systems evolves continuously as the one in larger systems and that a single mechanism sensitive to the system size could be responsible for nuclei production. The theoretical calculations qualitatively agree with the trend observed in data.

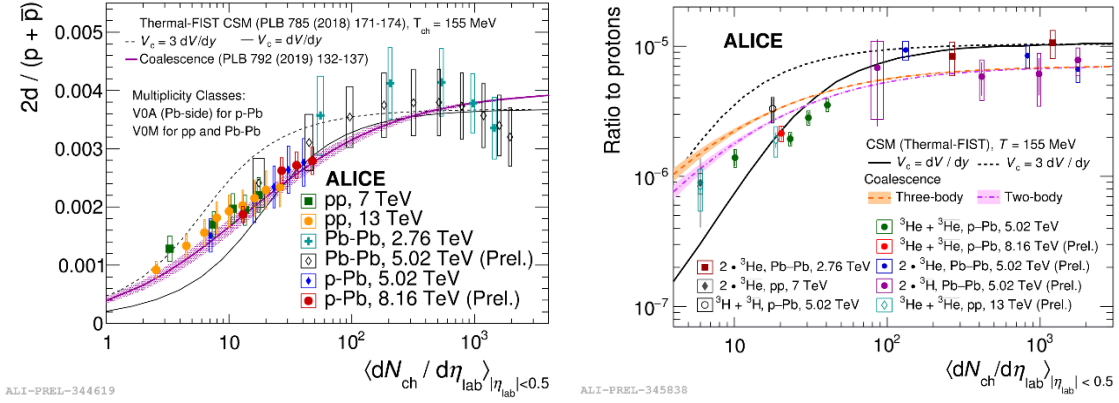

**Figure 2**: Deuteron (left panel), (anti)$^3$He and (anti)$^3$H (right panel) over proton ratios as a function of the charged-particle multiplicity in different collision systems and energies. Statistical uncertainties are represented as vertical lines whereas boxes represent the systematic ones. The results are compared to the expectations of SHM and Coalescence models.

### 4. Conclusions

The coalescence approach describes the experimental results concerning the ratio of the integrated yields of nuclei and protons as well as the coalescence parameter $B_A$ as a function of the charged-particle multiplicity density at midrapidity. For high charged-particle multiplicity densities, the coalescence approach and the CSM both succeed in the description of the d/p ratio, whereas models struggle to describe the ratio to protons for nuclei with A=3.

More precise measurements will be performed exploiting the data that will be collected during the ALICE Run 3 campaign and upcoming phenomenological calculations will help improving the theoretical understanding of the production mechanisms of (anti)nuclei in ultrarelativistic heavy-ion collisions.

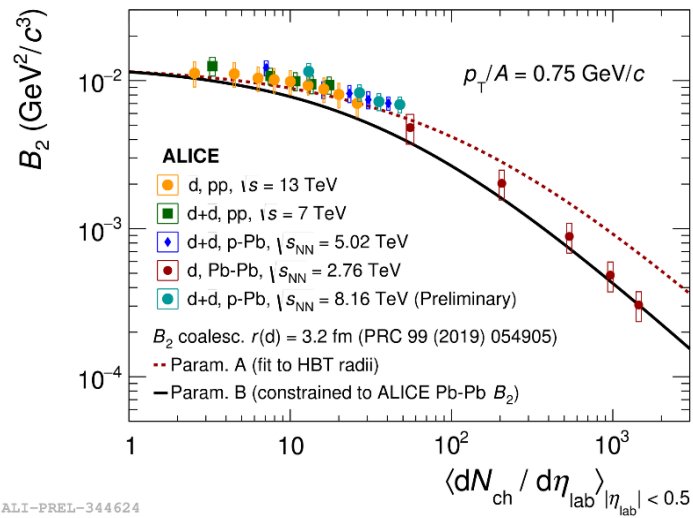

**Figure 3**: $B_2$ as a function of the mean charged-particle multiplicity density for a fixed value of $p_T$/A=0.75 GeV/c. The experimental results are compared to the coalescence calculations from [3] using two different parametrizations for the system size as a function of multiplicity.